\documentclass[prd,superscriptaddress,showpacs,amssymb,amsmath,amsfonts,aps,altaffilletter,nofootinbib,letterpaper,twocolumn]{revtex4-1}
\usepackage[caption=false]{subfig}
\usepackage{epsfig}
\usepackage{graphics}
\usepackage{graphicx}
\usepackage{amsmath,amssymb}
\usepackage{amsfonts}
\usepackage{bm}
\usepackage[usenames,dvipsnames]{color}
\usepackage{amssymb}
\usepackage{psfrag}
\usepackage{subfig}
\usepackage{tikz}
\usetikzlibrary{shapes,arrows}
\usepackage[usenames]{color}
\usepackage{multirow} % For the table
\usepackage[colorlinks]{hyperref}
\definecolor{LinkColor}{rgb}{0.741, 0.173, 0.000}
\definecolor{CiteColor}{rgb}{0.000, 0.502, 0.118}
\definecolor{UrlColor} {rgb}{0.000, 0.212, 0.800}
\hypersetup{linkcolor=LinkColor}
\hypersetup{citecolor=CiteColor}
\hypersetup{urlcolor=UrlColor}
\usepackage{url}
\usepackage{breakurl}

\newcommand{\msun}{M${}_{\odot}$}

\usepackage{ulem} 

\def\gw#1{gravitational wave#1 (GW#1)\gdef\gw{GW}}
\begin{document}

%%%%%%%%%%%%%%%%%%%%%%%%%%%%%%%%%%%%%%%%
\newcommand{\be}{\begin{equation}}
\newcommand{\ee}{\end{equation}}
\newcommand{\ber}{\begin{eqnarray}}
\newcommand{\eer}{\end{eqnarray}}
\newcommand{\bea}{\begin{eqnarray}}
\newcommand{\eea}{\end{eqnarray}}

%%%%%%%%%%%%%%%%%%%%%%%%%%%%%%%%%%%%%%%%

\title{Sensitivity Comparison of Searches for Binary Black Hole Coalescences with Ground-based Gravitational-Wave Detectors}
\author{%
Satya~Mohapatra$^{1,2,3}$,
Laura~Cadonati$^{1,4}$,
Sarah~Caudill$^{5,6}$,
James~Clark$^{1}$,
Chad~Hanna$^{7}$,
Sergey~Klimenko$^{8}$,
Chris~Pankow$^{5}$,
Ruslan~Vaulin$^{9}$,
Gabriele Vedovato$^{10}$,
and Salvatore~Vitale$^{9}$.
}\noaffiliation

\affiliation {University of Massachusetts Amherst, Amherst, MA 01003, USA }% {UMAmherst} {1}
\affiliation {Rochester Institute of Technology, 85 Lomb Memorial Drive, Rochester, NY 14623, USA}% {RIT} {2}
\affiliation {Syracuse University, Syracuse, NY 13244, USA }% {Syracuse} {3}
\affiliation {Cardiff University, Cardiff, CF24 3AA, United Kingdom }% {cardiff} {4}
\affiliation {University of Wisconsin--Milwaukee, Milwaukee, WI 53201, USA }% {UWM} {5}
\affiliation {Louisiana State University, LA 70803, USA}% {LSU} {6}
\affiliation {Pennsylvania State University, University Park, PA 16802, USA }% {PennState} {7}
\affiliation {University of Florida, Gainesville, FL 32611, USA }% {florida} {8}
\affiliation {LIGO, Massachusetts Institute of Technology, Cambridge, MA 02139, USA }% {MIT} {9}
\affiliation {INFN Sezione di Padova, 35131 Padova, Italy}% {INFN} {10}

\date{\today}

\begin{abstract}
\begin{center}

\end{center}

Searches for gravitational-wave transients from binary black hole coalescences typically rely on one of two approaches: matched filtering with templates and morphology-independent excess power searches.
Multiple algorithmic implementations in the analysis of data from the first generation of ground-based gravitational wave interferometers have used different strategies for the suppression of non-Gaussian noise transients, and  targeted different regions of the binary black hole parameter space.
In this paper we compare the sensitivity of three such algorithms: matched filtering with full coalescence templates,  matched filtering with ringdown templates and a morphology-independent excess power search.
The comparison is performed at a fixed false alarm rate and relies on Montecarlo simulations of binary black hole coalescences for spinning, non-precessing systems with total mass 25--350~\msun, which covers the parameter space of stellar mass and intermediate mass black hole binaries. 
We find that in the mass range of 25--100~\msun the sensitive distance of the search, marginalized over source parameters, is best with matched filtering to full waveform templates, to within 10\% at a false alarm rate of 3 events/year. In the mass range of 100-350~\msun, the same comparison favors the morphology-independent excess power search to within 20\%. The dependence on mass and spin is also explored. 

\end{abstract}

\maketitle

%%%%%%%%%%%%%%%%%%%%%%%%%%%%%%%%%%%%%%%%
\section{Introduction}
\label{sec:intro}
Binary black hole coalescences are amongst the most promising sources of gravitational-wave transients for ground based gravitational-wave observatories such as LIGO and Virgo~\cite{300years,ligo-ref,virgo-ref}. While stellar mass black holes with mass between 2.5~\msun{} and a few tens of \msun{} are formed by stellar collapse~\cite{polytrope,0004-637X-730-2-140,2041-8205-715-2-L138,0004-637X-759-1-52}, intermediate mass black holes between a few tens of \msun{} and $10^{5}$ \msun{} may result from the merger of stellar mass black holes or runaway collision of massive stars in dense globular clusters~\cite{imbh-globular-1,imbh-globular-2,0004-637X-734-2-111}. 
No evidence of binary black hole coalescence has been detected so far in data from initial LIGO and Virgo~\cite{inspiral,cbclowmass_s6,cbc-highmass,cbc-highmass-s6,S5_S6_ringdown,imbh,imbh-s6}. However, according to current rate predictions, advanced LIGO~\cite{adv-LIGO} and advanced Virgo~\cite{adv-Virgo} are expected to detect several gravitational-wave signals from binary black hole coalescences~\cite{rate}.

Following the seminal work of~\cite{flanhughes}, LIGO and Virgo data have been searched for separate phases of the binary black hole coalescence: inspiral, merger and ringdown~\cite{inspiral,cbclowmass_s6,imbh,ringdown}.
Search methods have evolved to account for full inspiral-merger-ringdown waveform templates~\cite{cbc-highmass}, use signal based vetoes~\cite{chi-square-bruce}, employ multi-resolution time-frequency information~\cite{multi-resolution}, and utilize coincident and coherent methods~\cite{robinson2008,ringdown,cwb}.

This paper introduces a framework to compare different searches in real detector noise, and applies it to algorithms representative of those used in recent searches for binary black holes in LIGO and Virgo data~\cite{cbc-highmass,cbc-highmass-s6,cwb,ringdown,S5_S6_ringdown,imbh-s6}. 
Section~\ref{sec:noise},  describes the data used in this study. Section~\ref{sec:algorithm} introduces the search algorithms that are compared in this study. Section~\ref{sec:method} outlines the method used in the a comparison, and Section~\ref{sec:search-performance} presents the results, using three complementary figures of merit.

%%%%%%%%%%%%%%%%%%%%%%%%%%%%%%%%%%%%%%%%
\section{Detector Data}
\label{sec:noise}
\begin{figure}[b]
\centering
\subfloat{\includegraphics[width=0.5\textwidth]{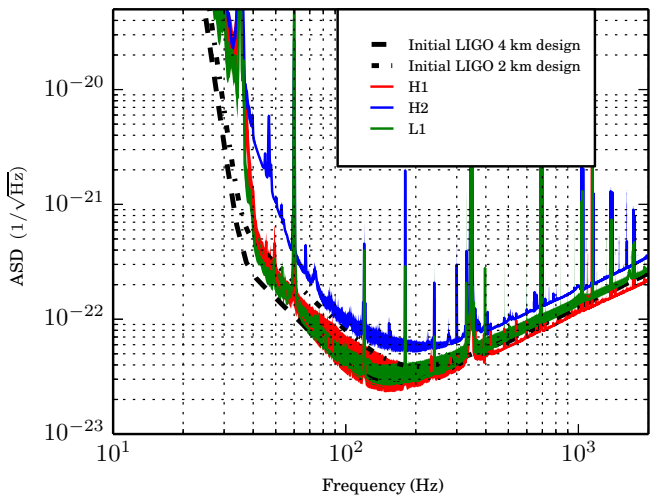}}
\caption{ Sensitivity of the detectors during the period of data in this study. 
The color shaded region indicates the $5^{\mathrm{th}}$ to $95^{\mathrm{th}}$ amplitude spectral density percentiles.} 
\label{fig:psd}
\end{figure}

We conducted this study on two months of data from the $5^{\mathrm{th}}$ science run LIGO  (14 Aug to 30 Sept 2007), when initial LIGO was at design sensitivity~\cite{s5-sensitivity}. We considered the three-detectors network of the 4-km and 2-km Hanford (H1, H2)~\cite{ligo-wa}, and 4-km Livingston (L1)~\cite{ligo-la} observatories.  
 Fig.~\ref{fig:psd} shows the detectors' sensitivity, expressed as strain amplitude spectral density (ASD). The color shaded region indicates the $5^{\mathrm{th}}$ to $95^{\mathrm{th}}$ ASD percentiles  in the analyzed period.

Data below 40 Hz is not included in this analysis, since it is limited by the seismic noise and, therefore, is not calibrated~\cite{calibration}. 
We applied data selections and vetoes to account for environmental artifacts and instrumental glitches that affect the quality and reliability of the data; this, combined with the instruments' duty cycle, resulted in a 3-detector lifetime of 29 days. See~\cite{s5-glitch,first_joint_ligo_geo_virgo_burst_s5} for details of this procedure.  

%%%%%%%%%%%%%%%%%%%%%%%%%%%%%%%%%%%%%%%%
\section{Search algorithms}
\label{sec:algorithm}
Transient gravitational-wave searches can be broadly classified into matched filtering with templates and unmodeled searches which do not assume a specific signal model. 
Matched filtering  can be performed with templates for the full binary black hole coalescence or only a portion of it. Unmodeled searches look for statistically significant excess signal energy in gravitational wave data. 
In this paper we consider three algorithms that have been adopted in searches of LIGO-Virgo data: the IMR-templates search~\cite{cbc-highmass,cbc-highmass-s6}, the ringdown search~\cite{ringdown,S5_S6_ringdown}, and coherent WaveBurst~\cite{imbh,imbh-s6}. 

\paragraph{IMR-templates ---}
\label{sec:highmass}
Matched filtering to a bank of non-spinning EOBNR templates~\cite{eobnr} has been used to search for binary black hole mergers in 25--100~\msun~\cite{cbc-highmass,cbc-highmass-s6}. Matched filtering is optimal for weak signals in Gaussian noise. However, due to the non-stationary, non-Gaussian nature of LIGO-Virgo data, this search is augmented with a sophisticated inter-site coincidence test~\cite{robinson2008}, a time-frequency, $\chi^{2}$ signal consistency test~\cite{chi-square-bruce} and a combined false alarm rate event ranking, all of which is described in Section 3 of~\cite{cbc-highmass}. 

\paragraph{Ringdown search ---}
\label{sec:ringdown}
The post-merger signal is accurately described as a superposition of quasi-normal oscillation modes, the {\it ringdown} waveforms~\cite{Teukolsky-1973ha}. Searches for ringdowns utilize a matched filtering algorithm with damped sinusoid templates characterized by quality factor $Q$ and central frequency $f_0$, parameters that describe the $\ell=m=2$ oscillation mode of the final merged black hole~\cite{creighton-ring-template,ringdown,S5_S6_ringdown}. 
Details on the template bank and the algorithm are given in~\cite{S5_S6_ringdown}. The central frequency and quality factor can be empirically related to the final mass and spin of the merged black hole~\cite{berti_2006}; the search space corresponds roughly to black holes with masses in the range 10 \msun{} to 600 \msun{} and spins in the range 0 to 0.99.

The ringdown matched filter search is also augmented with a sophisticated multi-detector coincidence test~\cite{robinson2008,nakano2003}, an amplitude consistency test~\cite{S5_S6_ringdown}, and an event ranking based on the {\it chopped-L statistic} described in~\cite{talukder2003}. This event ranking statistic differs from the multivariate statistic used in~\cite{S5_S6_ringdown}. Candidate events are ranked 
with the combined false alarm rate detection statistic described in~\cite{cbc-highmass,S5_S6_ringdown}.

\paragraph{Coherent WaveBurst ---}
\label{sec:cwb}
The coherent WaveBurst (CWB) algorithm is designed to identify coherent excess power transients without prior knowledge of the waveform, and it is used in searches for gravitational-wave bursts in LIGO and Virgo data~\cite{first_joint_ligo_geo_virgo_burst_s5,second_joint_ligo_virgo_burst_s6}.
By imposing weak model constraints, such as the requirement of elliptical polarization, CWB has been optimized 
to search for black hole binaries of total masses between 100 -- 450 \msun~\cite{imbh,imbh-s6}. 

The algorithm uses a wavelet basis to decompose the data into a time-frequency map with discrete signal energy pixels. 
The algorithm then executes a constrained maximum likelihood analysis of the decomposed network data stream. Reconstruction of detector responses occurs for a trial set of sky locations and corresponding arrival delays. The residual data, after subtraction of the estimated detector responses from the original data, represents the reconstructed network noise. 
The elliptical polarization constraint
is expected to have minimal impact on the recovery of gravitational-wave signals from compact binary coalescences, while enhancing the rejection of noisy events~\cite{elliptical-polarization}. 

The coherent network amplitude $\eta$ is the detection statistics used by CWB~\cite{imbh,imbh-s6}. It is proportional to the average signal to noise ratio (SNR) per detector and is used to rank selected events and establish their significance against a sample of background events.

%%%%%%%%%%%%%%%%%%%%%%%%%%%%%%%%%%%%%%%%
\section{Comparison Method}
\label{sec:method}
Since the algorithms described in section~\ref{sec:algorithm} have different targets, it is natural to expect they respond differently to weak signals in the data and to noise transients.

In this paper we compare the sensitivity of the three searches via their false alarm rate, a search independent ranking of an event's rate of occurrence, as determined from a background sample.
We ran the analyses with configurations  that are representative of their applications in published results~\cite{cbc-highmass,cbc-highmass-s6,ringdown,S5_S6_ringdown,imbh,imbh-s6}.

For this analysis, we injected in the detectors' noise simulated gravitational-wave signals from the coalescence of binary black holes.
The waveforms were produced with the IMRPhenomB~\cite{spin-phenom} model: a phenomenological, non-precessing, spinning binary black hole template family, which tracks the coalescence from late inspiral to ringdown. 
 In this configuration, the spin vectors ($\chi_{1}$ and $\chi_{2}$) are aligned/anti-aligned with the angular momentum of the binary system. The waveforms are parametrized by three physical parameters: the component black hole masses $m_{1}$, $m_{2}$, and the mass weighted spin parameter $\chi_{s}$,

\begin{equation}
\chi_{s}=\frac{m_{1} \chi_{1} + m_{2} \chi_{2}}{m_{1}+m_{2}}.
\label{chi_s}
\end{equation}

The waveforms do not include the effects of non-aligned spin-orbit coupling, but do account for aligned / anti-aligned spin-orbit interaction, such as the orbital hang-up effect~\cite{orbital-hangup}.

 To determine the false alarm rate, we time shifted the data from one or more detectors well beyond the light travel time of 10 ms between H1 and L1. We imposed a minimum shift of 5 seconds to remove inter-site correlations which could be due to a real gravitational-wave signal in the data. We did not introduce time shifts between data from the co-located H1 and H2 detectors, since the background should account for site-specific correlations~\cite{cbc-highmass}.
We applied 100 equally spaced time shift in the ringdown and the IMR-templates searches, and 600 in the CWB search. 
We declared an injection {\em detected} if a coincident event was identified within 100 ms of the nominal injection time. This interval is long enough to account for the uncertainty in identifying the {\em arrival time} of a signal, where the arrival time is the maximum amplitude of the waveform.

  Each pipeline ranked all the events and assigned a false alarm rate by comparison with its native background ranking statistics.

We evaluated detection efficiency and sensitive distance (as defined in section~\ref{sec:search-performance}) for a range of measurable false alarm rate thresholds. We quote the results for a false alarm rate threshold of 3 events per year which is in the middle of this range.  We made sure that the searches use consistent data after the application of data quality vetoes, with small differences due to technical details in the veto implementation~\cite{s5-glitch}.

%%%%%%%%%%%%%%%%%%%%%%%%%%%%%%%%%%%%%%%%
\section{Results}
\label{sec:search-performance}
\subsection{Target parameter space}

In this study, we partitioned the parameter space according to the total mass of the binary system.
\textit{Set A} includes systems with total mass between 25 and 100 \msun, as searched by IMR-templates~\cite{cbc-highmass,cbc-highmass-s6}. \textit{Set B} consists of total mass between 100 and 350 \msun, which overlaps the parameter space searched by the CWB algorithm~\cite{imbh,imbh-s6}.
We restricted the total mass to below 350 \msun{} as the peak detectable frequencies from the ringdown for some of the spin configuration is below 40 Hz for mass above 350 \msun~\cite{spin-phenom,final-spin}, and thus is subject to unacceptable or ill-defined uncertainties arising from  calibration~\cite{calibration}. 

Simulated signals are uniformly  distributed   in total mass ($m$), mass ratio ($q$), and dimensionless spin parameter $\chi_{s}$, in the intervals listed in  Table~\ref{tbl:inj_param}.
This distribution is not meant to reproduce the expected astrophysical distribution of binary black hole sources, but rather to probe a wide physical parameter space and evaluate the efficacy of each pipeline in detection. The injections are logarithmically distributed in distance.  No correction to the waveform due to redshift at cosmological distances is included, as this effect is expected to be small ($z$ $<=$ 0.1) at the reach of initial detectors. 
Injections are also uniformly distributed in sky location, polarization and inclination of the binary relative to Earth.

We analyzed $\sim$25000 
injections; due to the limitations of the search (i.e. reduced efficacy of $\chi^{2}$ above 100 \msun{} as discussed in section~\ref{sec:algorithm}), we used  
the IMR-templates search only for the lower mass set of injections. We performed ringdown matched filter and CWB analyses on both injection sets.

\begin{table}[htbp]
\begin{center}
\caption{Simulated waveform parameters.}
\begin{tabular}{lr}
\hline
Total Mass (\msun), $m$: Set A & 25 -- 100 \\
Total Mass (\msun), $m$: Set B & 100 -- 350  \\
Mass Ratio (both sets), $q$: & 0.1 -- 1  \\ 
$\chi_{s}$ (both sets) & -0.85 -- 0.85 \\
Distance (Mpc) (both sets) & 0 -- 2000 \\
\hline
\end{tabular}
\label{tbl:inj_param}
\end{center}
\end{table}

\begin{figure}[h]
\centering
\subfloat{\includegraphics[width=0.4\textwidth]{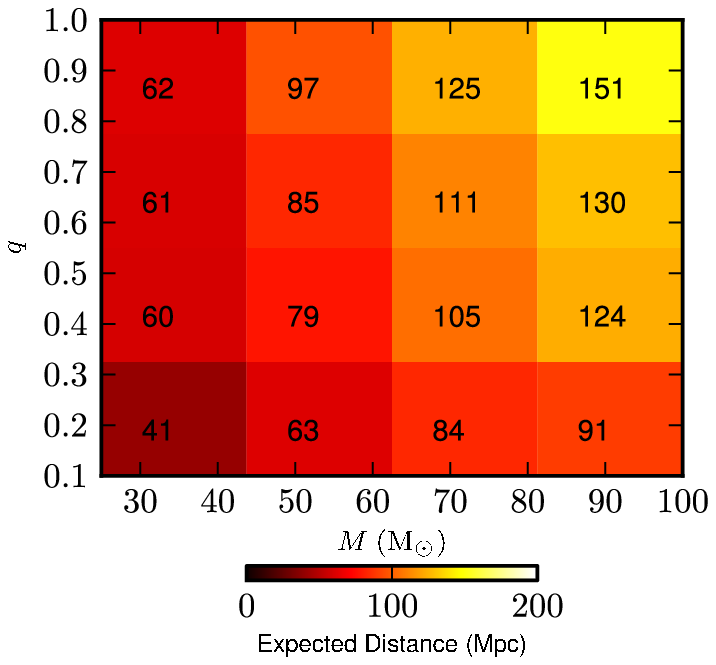} \label{fig:m_q-average-range-IMR-a}}\\
\subfloat{\includegraphics[width=0.4\textwidth]{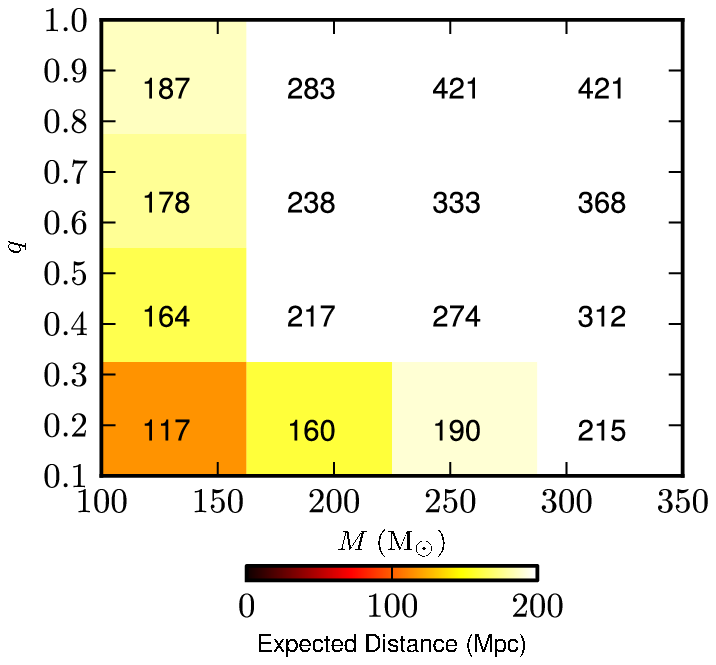} \label{fig:m_q-average-range-IMR-b}}\\
\caption{Expected range (Mpc) as a function of $m$ vs $q$. This quantity is estimated from the network SNR threshold of 12, and is marginalized over the spin parameter, sky position and orientation of the binary system. The color scale is saturated at 200 Mpc for comparison with Figs.~\ref{fig:sensitive-distance-IMR-setA-m-q},~\ref{fig:sensitive-distance-IMR-setA-m-chi},~\ref{fig:sensitive-distance-IMR-setB-m-q},~\ref{fig:sensitive-distance-IMR-setB-m-chi}.}
\label{fig:m_q-average-range}
\end{figure}

\begin{figure}[h]
\centering
{\includegraphics[width=0.4\textwidth]{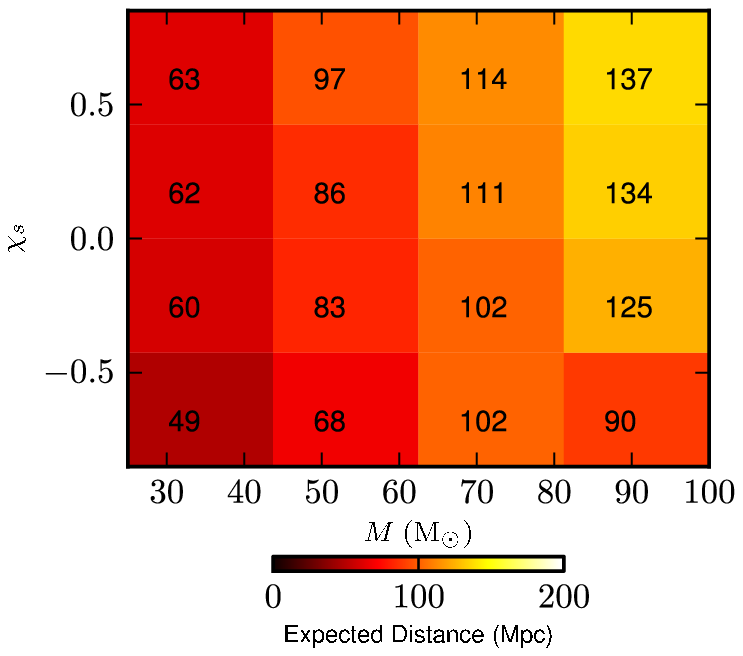} \label{fig:m_chi-average-range-IMR-a}}\\
{\includegraphics[width=0.4\textwidth]{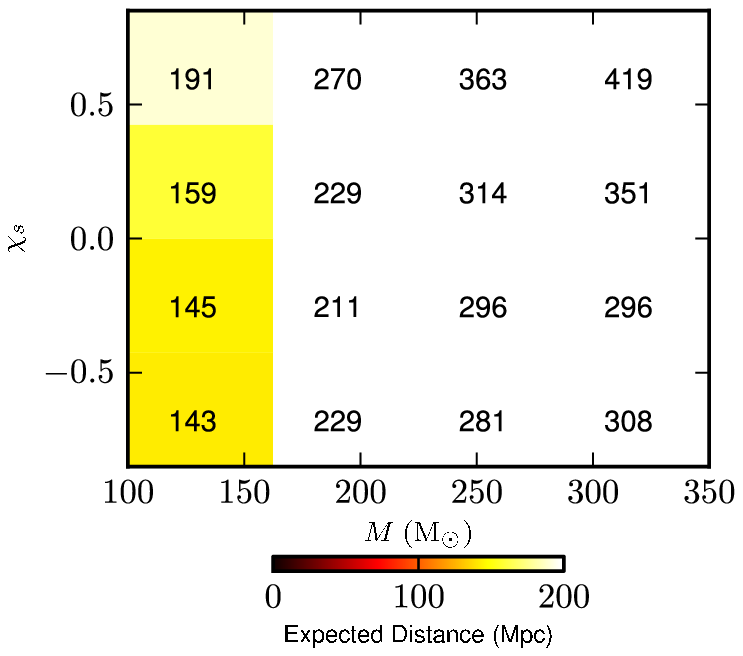} \label{fig:m_chi-average-range-IMR-b}}\\
\caption{Expected range (Mpc) as a function of $m$ $\chi$. This quantity is estimated at a network SNR threshold of 12, and is marginalized over mass ratio, sky position and orientation of the  black hole binary system.}
\label{fig:m_chi-average-range}
\end{figure}

Fig.~\ref{fig:m_q-average-range} and~\ref{fig:m_chi-average-range} show the \emph{expected range} as a function of  total mass, mass ratio and spin parameter. The expected range is calculated by averaging the distances over extrinsic parameters  such as sky position and inclination of the binary black holes for which the network signal-to-noise ratio (SNR) is 12, following the prescription used in~\cite{obs-scenario}.
The SNR is estimated from the median value of the amplitude spectral density of the instrumental noise in Fig.~\ref{fig:psd}.

Fig.~\ref{fig:m_q-average-range} illustrates that 
the expected range is higher for symmetric mass binary black holes compared to asymmetric mass system for the same total mass.  
This is consistent with the fact that the SNR of the signal is proportional to its amplitude  divided by the square-root of its duration in time. For a binary black hole, the gravitational-wave amplitude is proportional to $\dfrac{q}{(1+q)^2}$, while the time duration is proportional to $\dfrac{(1+q)^{2}}{q}$, hence SNR is proportional to $\dfrac{\sqrt{q}}{1+q}$~\cite{0264-9381-24-12-S04}.
Fig.~\ref{fig:m_chi-average-range} illustrates that the expected range is higher for aligned than anti-aligned spin configurations, since systems with aligned spins stay longer in orbit until merger, hence get more relativistic, leading to higher gravitational-wave luminosity, than  anti-aligned systems~\cite{orbital-hangup}.

\subsection{Search performance}

We now define the quantities that will be used for the comparison.
The detection efficiency $\varepsilon$ of a search is a function of the false alarm rate threshold 
$\zeta$,  the radial distance to the source $r$, the total mass $m$, the mass ratio $q$, and the spin parameter $\chi_{s}$. 
In this work, we average over sky location, polarization and orientation, and define the average efficiency as:

\begin{equation}
\bar{\varepsilon}(\zeta,r,m,q,\chi_{s}) = \frac{N_{f}}{N_{i}},
\label{eff-eqn}
\end{equation}
where $N_{f}$ is the number of found injections and $N_{i}$ is the number of total injections averaged over all sky position and inclination.

The sensitive volume, or the volume of the sky surveyed is defined as:
\begin{equation}
V(\zeta,m,q,\chi_{s}) = \int 4 \pi r^{2} \bar{\varepsilon}\,dr. 
\label{sensvol-eqn}
\end{equation}

Finally, the sensitive radius $\mathcal{R}$ is the radius of the sphere with volume of $V$: 
\begin{equation}
\mathcal{R}(\zeta,m,q,\chi_{s})   = \left[ \frac{3}{4\pi}V
\right] ^{1/3}.
\label{sensdist-eqn}
\end{equation}

\paragraph*{Efficiency curves ---}
\label{sec:sig}

\begin{figure}
\centering
\subfloat[Part 1][Total mass 25--100 \msun.]{\includegraphics[width=0.4\textwidth]{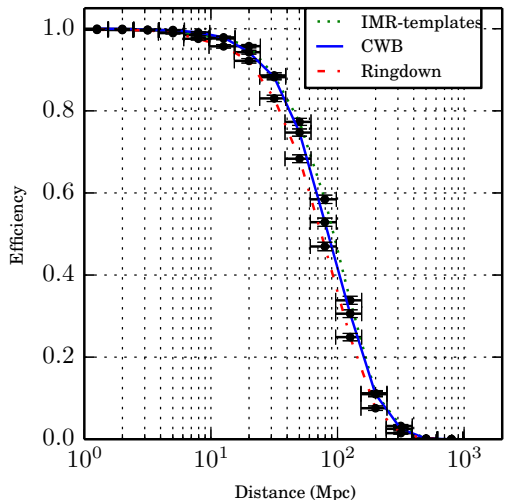} \label{fig:sigmoid-IMR-algos-a}}\\
\subfloat[Part 3][Total mass 100--350 \msun.]{\includegraphics[width=0.4\textwidth]{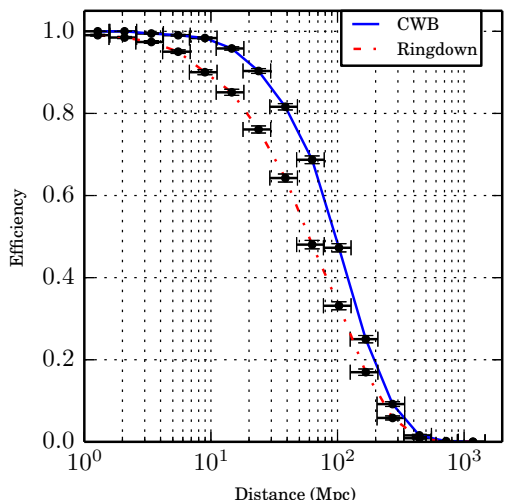} \label{fig:sigmoid-IMR-algos-c}}

\caption{Efficiency curves at a false alarm rate of 3 events per year,  averaged over total mass, mass ratio, spin parameter, sky-position and inclination of the source.}
\label{fig:sigmoid-IMR-algos}
\end{figure}

The detection efficiency at fixed false alarm rate and distance is estimated from Eqn.~\ref{eff-eqn}; we plot it as a function of distance in Fig.~\ref{fig:sigmoid-IMR-algos},  where $N_i$ and $N_f$ are marginalized over all other source parameters. The horizontal and the vertical error bars in Fig.~\ref{fig:sigmoid-IMR-algos} are set by the bin boundaries and binomial statistics on the number of injected signals in each amplitude bin, respectively.

For a quantitative comparison we fit a cubic spline to the efficiency curve. We then compare two characteristic parameters, the 50\% and 90\% efficiency distances, $D_{\mathrm{eff}}^{50\%}$ and $D_{\mathrm{eff}}^{90\%}$, which are the distances at which 50\% and 90\% of the signals can be found, respectively in Table~\ref{tab:compare_searches_eff_at_far}. 

In the 25--100 \msun{} range (set A) the IMR-templates search yields 12\% and 25\% higher $D_{\mathrm{eff}}^{50\%}$, and 7\% and 30\% higher $D_{\mathrm{eff}}^{90\%}$ compared to  the CWB and the ringdown searches, respectively. In the 100--350 \msun{} range (set B), CWB search 60\% higher $D_{\mathrm{eff}}^{50\%}$ and 170\% higher $D_{\mathrm{eff}}^{90\%}$ compared to the ringdown search.

%%%%%%%%%%%%%%%%%%%%%%%%%%%%%%%%%%%%%%%%%%%%%%%%%
\begin{table}[hb]
\begin{center}
\caption{Efficiency distances at a false alarm rate of 3 events per year. }
\label{tab:compare_searches_eff_at_far}
\begin{tabular}{ccc}
  \colrule
  Algorithm (on Set A) & $D^{50\%}_{\mathrm{eff}}$  (Mpc) & $D^{90\%}_{\mathrm{eff}}$  (Mpc) \\
  \colrule
IMR-templates search   & 94 & 30 \\
CWB   & 84 & 28 \\
Ringdown   & 75 & 23 \\
%%  \colrule
\end{tabular}
\begin{tabular}{ccc}
  \colrule
  Algorithm (on Set B) & $D^{50\%}_{\mathrm{eff}}$  (Mpc) & $D^{90\%}_{\mathrm{eff}}$  (Mpc) \\
  \colrule
CWB   & 97 & 24 \\
Ringdown   & 60 & 9 \\
  \colrule
\end{tabular}
\end{center}
\end{table}

\begin{figure}[h]
\centering
\subfloat[Part 1][Total mass 25--100 \msun.]{\includegraphics[width=0.4\textwidth]{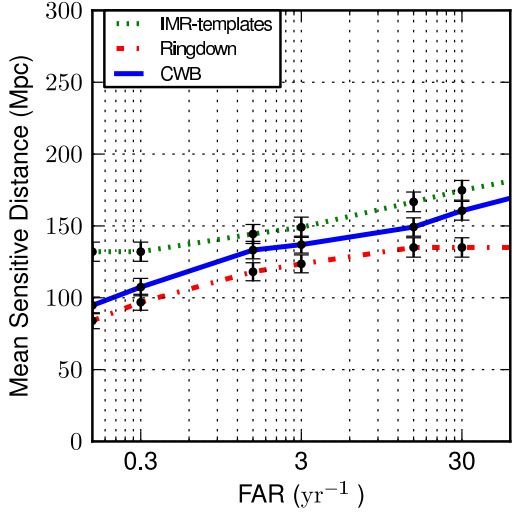} \label{fig:mean-sensitivity-distance-IMR-a}}\\
\subfloat[Part 2][Total mass 100--350 \msun.]{\includegraphics[width=0.4\textwidth]{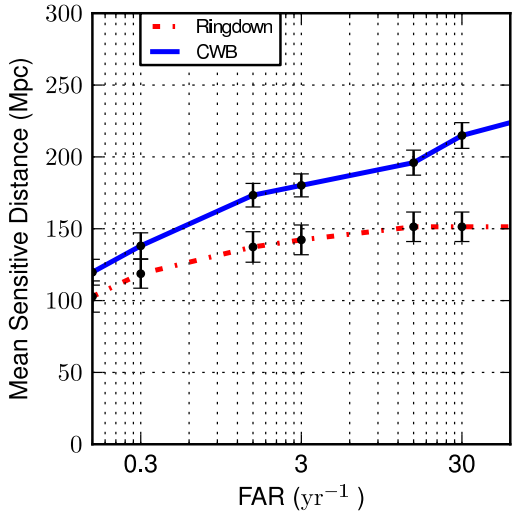} \label{fig:mean-sensitivity-distance-IMR-b}}

\caption{Mean sensitive distance as a function of false alarm rate.}
\label{fig:mean-sensitive-distance-IMR}
\end{figure}

\paragraph*{Mean sensitive distance ---}
\label{sec:roc}
Eqn.~\ref{sensdist-eqn} can be marginalized over all  parameters, to compute a mean sensitive distance as a function of the false alarm rate. This quantity is show in Fig.~\ref{fig:mean-sensitive-distance-IMR}.

We notice that within a range of false alarm rates of 0.3 events per year to 30 events per year, the three searches give consistent results. We do not see any abrupt change of sensitivity for a search over this range of false alarm rates. The false alarm rate of 3 events/year we chose to plot efficiency curves and quote sensitive distances is thus representative of the relative performance of the algorithms. 

Published searches have typically chosen lower FAR thresholds estimated on the loudest events~\cite{loudest1,loudest2} seen by the searches in open-box data: 0.2 events/year and 0.41 events/year  for IMR-templates searches on 2005-2007~\cite{cbc-highmass} and 2009-2010~\cite{cbc-highmass-s6} LIGO-Virgo data, 0.45 events/year for the ringdown search~\cite{S5_S6_ringdown}, and 0.76 events/year for the CWB search on 2005-2007 data~\cite{imbh}.

%%%%%%%%%%%%%%%%%%%%%%%%%%%%%%%%%%%%

\paragraph*{Sensitive distance ---}
\label{sec:sensitive}
To probe how different algorithms respond to different regions of the parameter space, we plot the sensitive distance, defined in Eqn.~\ref{sensdist-eqn}, as a function of the mass and spin parameters at a false alarm rate of 3 events per year. Additional sensitive distance plots in between 0.3 to 30 events per year are available at~\cite{science-summary}.

\begin{figure}
\centering
\subfloat[Part 1][Matched filter to IMR-templates.]{\includegraphics[width=0.4\textwidth]{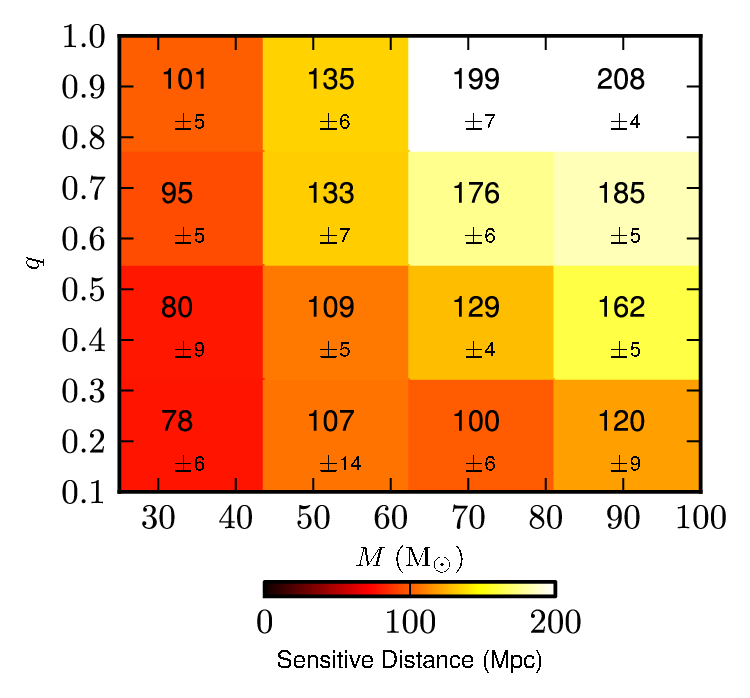} \label{fig:m_q-sensitivity-distance-IMR-a}}\\
\subfloat[Part 2][Coherent WaveBurst template-less search.]{\includegraphics[width=0.4\textwidth]{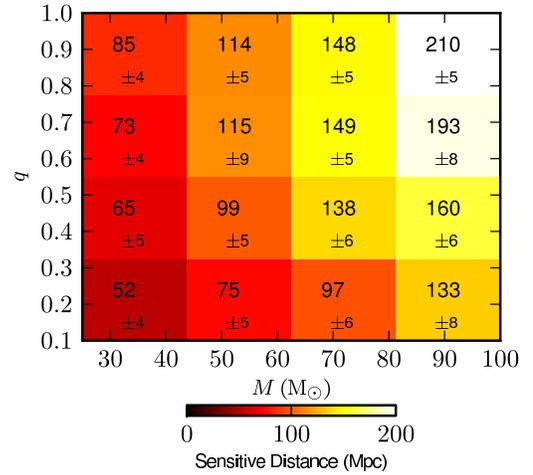} \label{fig:m_q-sensitivity-distance-IMR-b}}\\
\subfloat[Part 3][Matched filter to ringdowns.]{\includegraphics[width=0.4\textwidth]{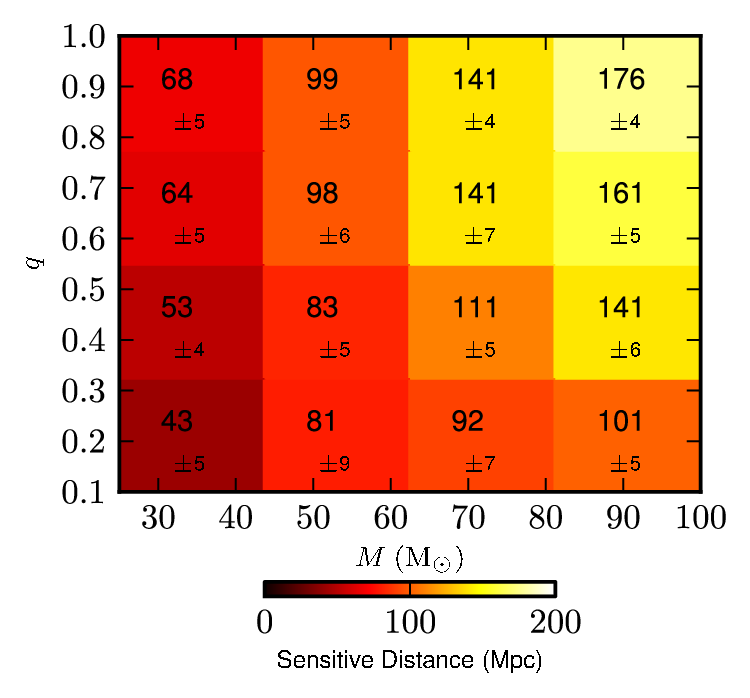} \label{fig:m_q-sensitivity-distance-IMR-c}}\\
\caption{Sensitive distance for systems with total mass 25--100 \msun{} as a function of total mass $m$ and  mass ratio $q$ at a false alarm rate of 3 events per year.}
\label{fig:sensitive-distance-IMR-setA-m-q}
\end{figure}

\begin{figure}
\centering
\subfloat[Part 7][Matched filter to IMR-templates.]{\includegraphics[width=0.4\textwidth]{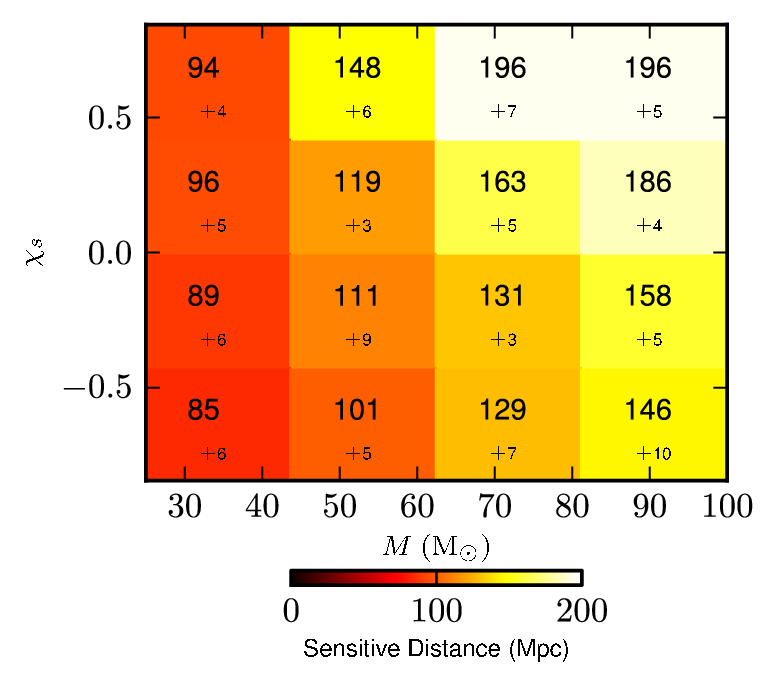} \label{fig:m_q-sensitivity-distance-IMR-f}}\\
\subfloat[Part 8][Coherent WaveBurst template-less search.]{\includegraphics[width=0.4\textwidth]{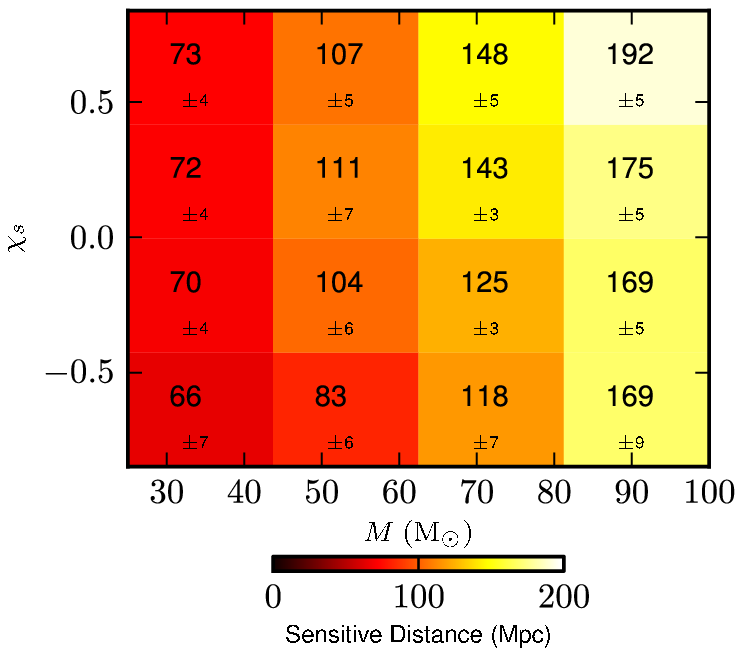} \label{fig:m_q-sensitivity-distance-IMR-e}}\\
\subfloat[Part 9][Matched filter to ringdowns.]{\includegraphics[width=0.4\textwidth]{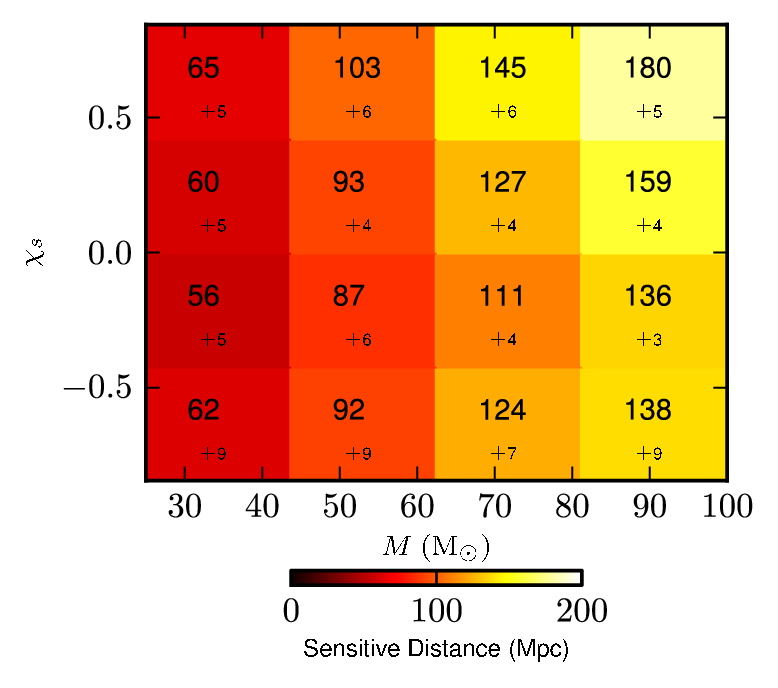} \label{fig:m_q-sensitivity-distance-IMR-g}}\\

\caption{Sensitive distance for systems with total mass 25--100 \msun{} as a function of total mass $m$ and  mass ratio $\chi_{s}$ at a false alarm rate of 3 events per year.}
\label{fig:sensitive-distance-IMR-setA-m-chi}
\end{figure}

\begin{figure}
\centering
\subfloat[Part 4][IMR-templates and Coherent WaveBurst.]{\includegraphics[width=0.4\textwidth]{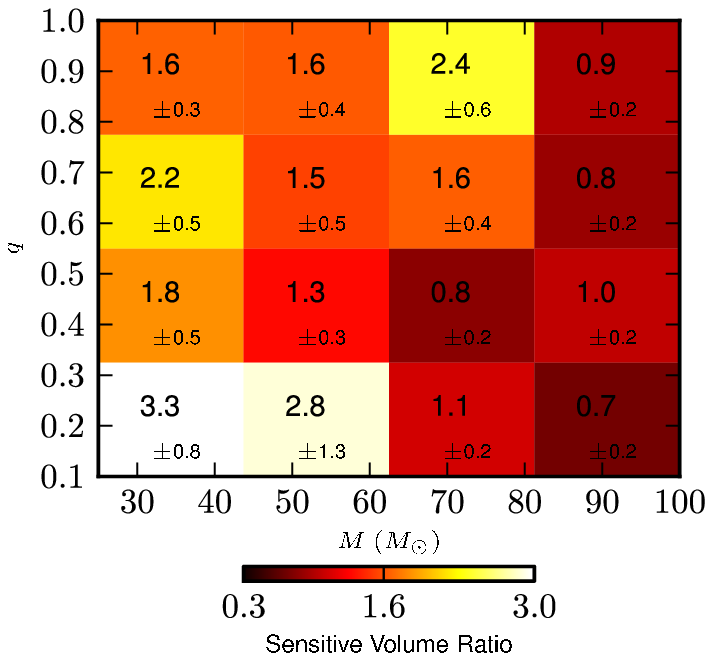} \label{fig:m_q-sensitivity-distance-IMR-d}}\\
\subfloat[Part 5][IMR-templates and Ringdown.]{\includegraphics[width=0.4\textwidth]{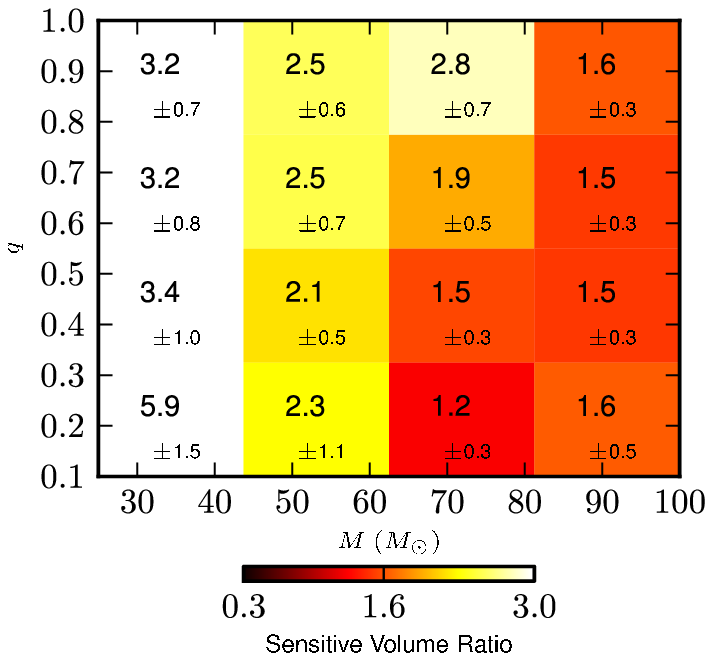} \label{fig:m_q-sensitivity-distance-IMR-b}}\\

\caption{Sensitive volume ratio for systems with total mass 25--100 \msun{} as a function of $m$ and  $q$ at a false alarm rate of 3 events per year.}
\label{fig:sensitive-distance-difference-IMR-setA-m-q}
\end{figure}

\begin{figure}
\centering
\subfloat[Part 10][IMR-templates and Coherent WaveBurst.]{\includegraphics[width=0.4\textwidth]{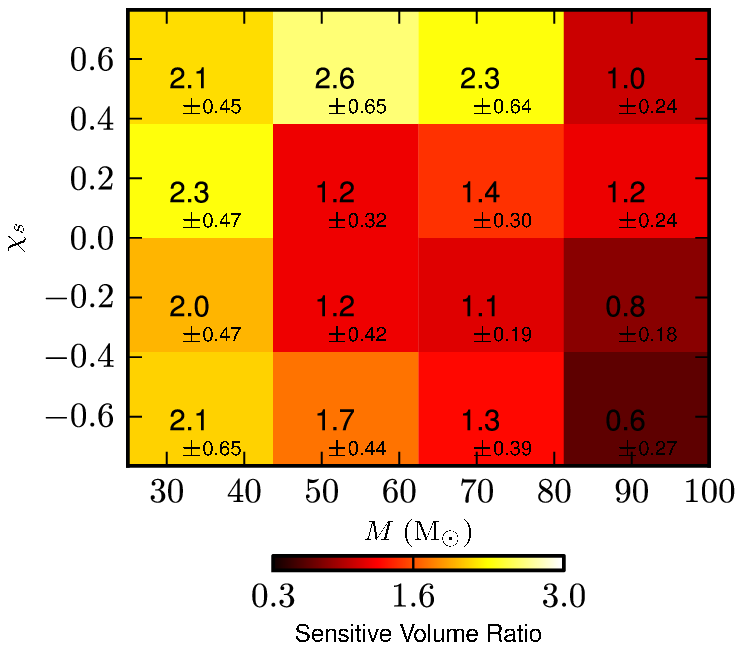} \label{fig:m_q-sensitivity-distance-IMR-h}}\\
\subfloat[Part 11][IMR-templates and Ringdown.]{\includegraphics[width=0.4\textwidth]{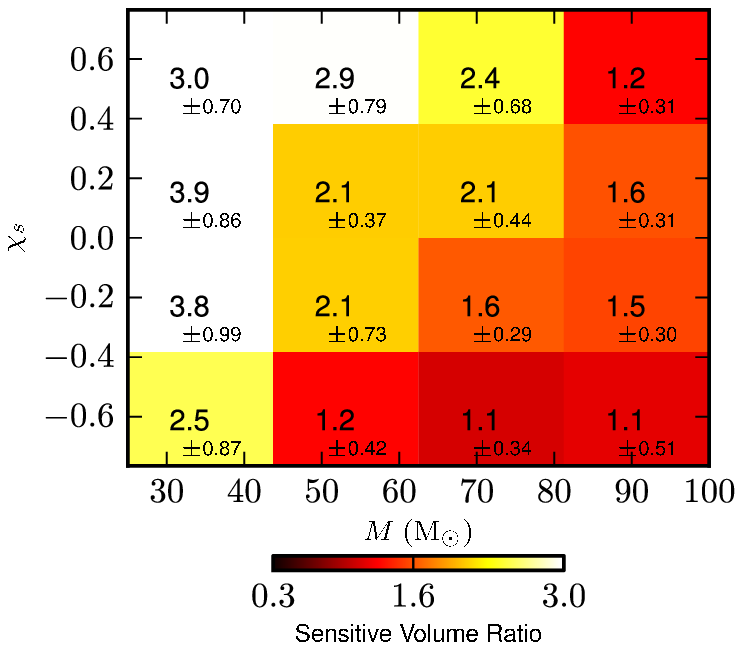} \label{fig:m_q-sensitivity-distance-IMR-i}}
\caption{Sensitive volume ratio for systems with total mass 25--100 \msun{} as a function of $m$ vs $\chi_{s}$ at a false alarm rate of 3 events per year.}
\label{fig:sensitive-distance-difference-IMR-setA-m-chi}
\end{figure}

\begin{figure}[h]
\centering
\subfloat[Part 1][Coherent WaveBurst template-less search.]{\includegraphics[width=0.4\textwidth]{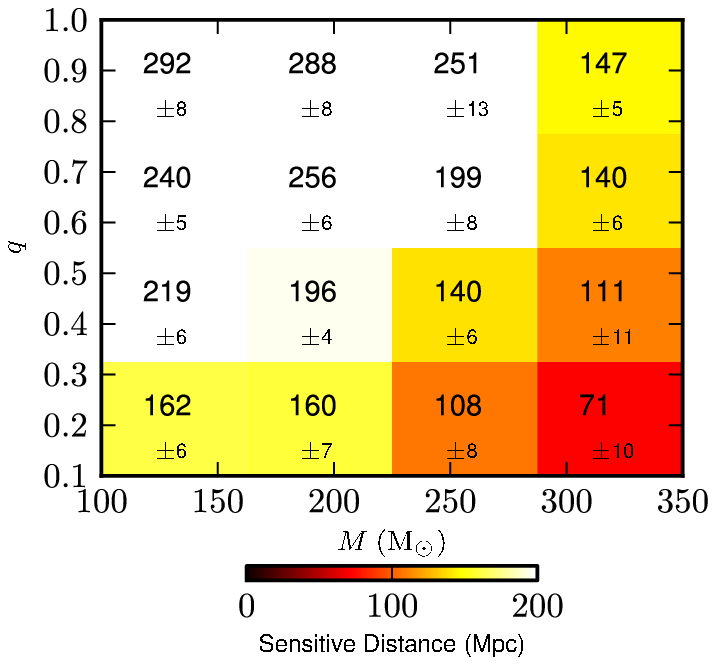} \label{fig:m_q-sensitivity-distance-IMR-k}}\\
\subfloat[Part 2][Matched filter to ringdowns.]{\includegraphics[width=0.4\textwidth]{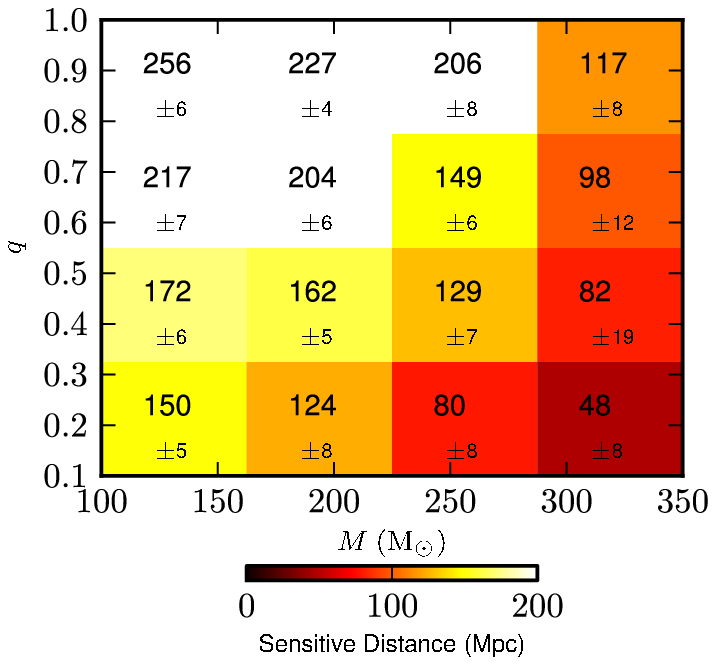} \label{fig:m_q-sensitivity-distance-IMR-l}}\\

\caption{Sensitive distance for systems with total mass 100--350 \msun{} as a function of $m$ and $q$ at a false alarm rate of 3 events per year.}
\label{fig:sensitive-distance-IMR-setB-m-q}
\end{figure}

\begin{figure}[h]
\centering
\subfloat[Part 1][Coherent WaveBurst template-less search.]{\includegraphics[width=0.4\textwidth]{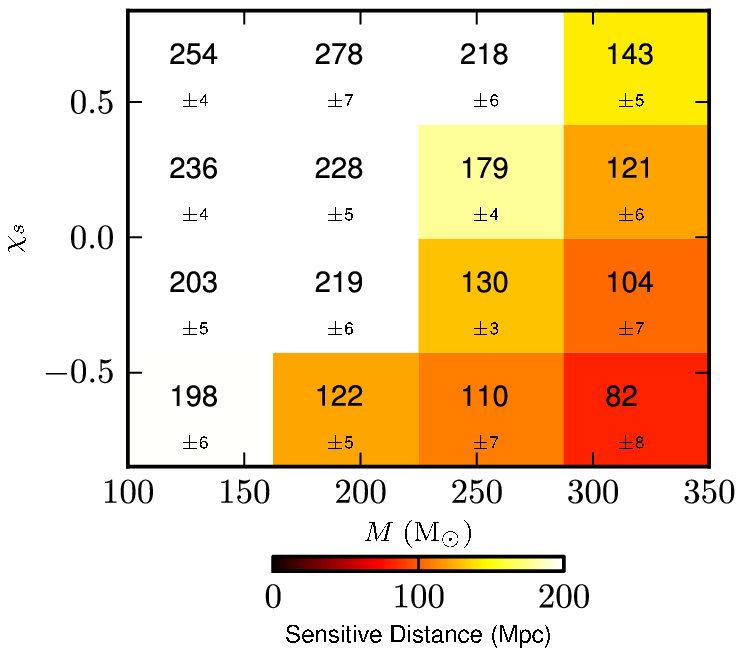} \label{fig:m_q-sensitivity-distance-IMR-n}}\\
\subfloat[Part 2][Matched filter to ringdowns.]{\includegraphics[width=0.4\textwidth]{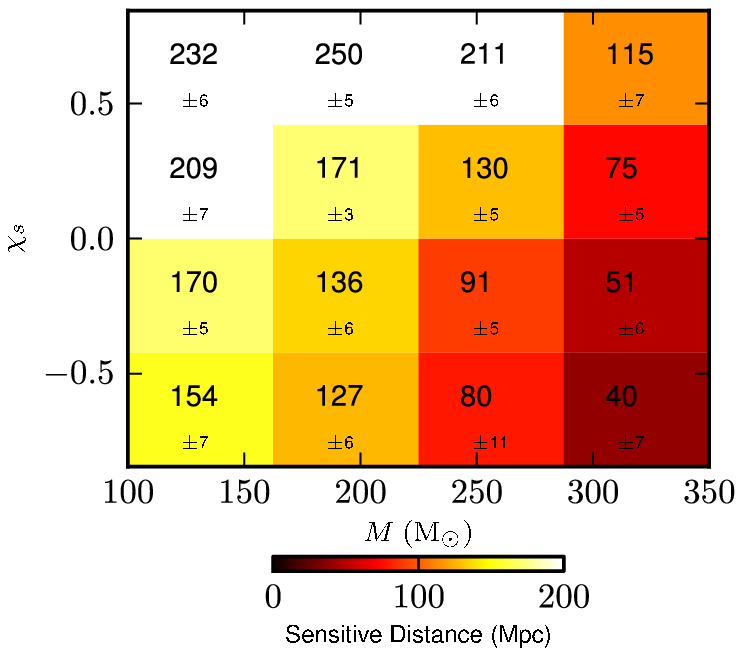} \label{fig:m_q-sensitivity-distance-IMR-o}}\\

\caption{Sensitive distance for systems with total mass 100--350 \msun{} as a function of $m$ and $\chi_{s}$ at a false alarm rate of 3 events per year.}
\label{fig:sensitive-distance-IMR-setB-m-chi}
\end{figure}

Fig.~\ref{fig:sensitive-distance-IMR-setA-m-q} and~\ref{fig:sensitive-distance-IMR-setB-m-q} show the sensitive distance at FAR threshold of 3 events per year as a function of mass parameters, $m$ and $q$, and are marginalized over the spin parameter $\chi_{s}$. Across the mass ranges,  
the three search algorithms are more sensitive to symmetric than asymmetric binary systems. 

For set A all the three search algorithms have the highest sensitive distance in the total mass bin of 80 to 100 \msun. For set B the CWB search and the ringdown search register higher sensitive distance in the total mass bin of 100 to 150 \msun.

Fig.~\ref{fig:sensitive-distance-IMR-setA-m-chi} and~\ref{fig:sensitive-distance-IMR-setB-m-chi} show the sensitive distance as a function of total mass and the spin parameter, and are marginalized over mass ratio, $q$.
Across the mass ranges, the three search algorithms have higher sensitivity for detecting aligned (with respected to the orbital angular momentum) spin binary black holes compared to the anti-aligned spin binary black holes~\cite{cbc-highmass-s6,imbh-s6,ninja-mdc}. Qualitatively, this observation is in consistent with the expected range in Fig.~\ref{fig:m_chi-average-range}. Quantitatively, sensitive distance differ from the expected range, which motivates the choice of false alarm rate as the criteria for detectability, rather than the signal SNR.

The errors quoted in Fig.~\ref{fig:sensitive-distance-IMR-setA-m-q},~\ref{fig:sensitive-distance-IMR-setA-m-chi},~\ref{fig:sensitive-distance-IMR-setB-m-q} and ~\ref{fig:sensitive-distance-IMR-setB-m-chi} are derived from the binomial statistics of events in each bin. 

\begin{figure}
\centering
\subfloat[Part 3][Coherent WaveBurst and Ringdown.]{\includegraphics[width=0.4\textwidth]{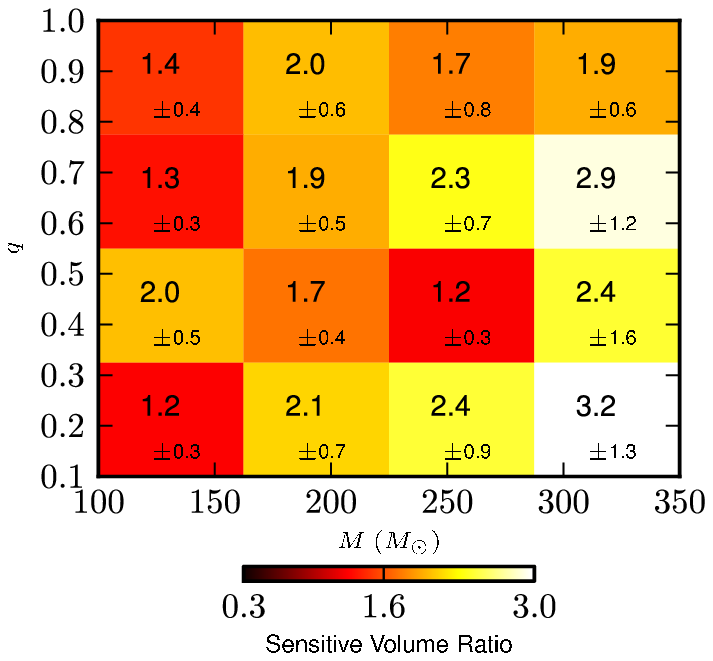} \label{fig:m_q-sensitivity-distance-IMR-m}}\\
\subfloat[Part 6][Coherent WaveBurst and Ringdown.]{\includegraphics[width=0.4\textwidth]{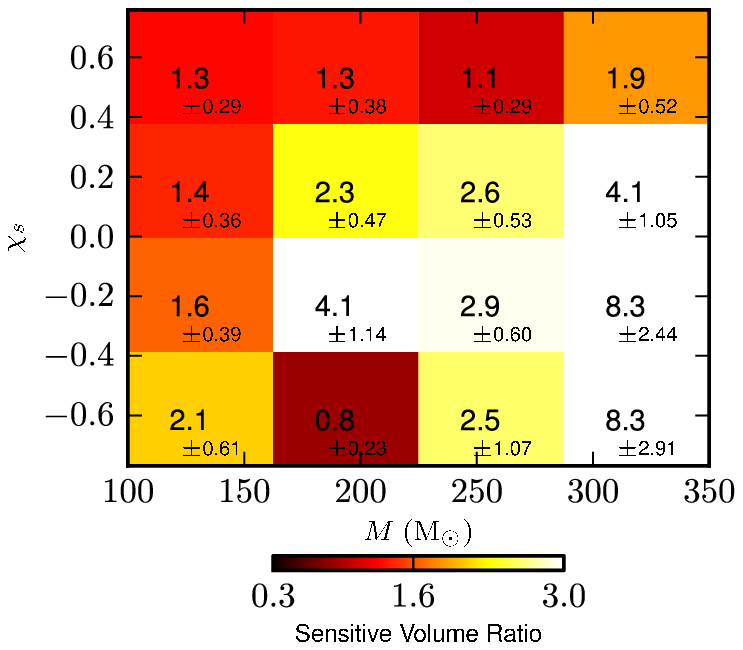} \label{fig:m_q-sensitivity-distance-IMR-p}}

\caption{Sensitive volume ratio for systems with total mass 100--350 \msun{} as a function of $m$ and $q$, and $m$ and $\chi_{s}$ at a false alarm rate of 3 events per year.}
\label{fig:sensitive-distance-difference-IMR-setB}
\end{figure}

Fig.~\ref{fig:sensitive-distance-difference-IMR-setA-m-q} shows the ratio in the sensitive volume, defined in Eq~\ref{sensvol-eqn}, for Set A injections in total mass and mass ratio, and Fig.~\ref{fig:sensitive-distance-difference-IMR-setA-m-chi} shows that ratio for total mass and spin parameter space. In these plots the significance, \textit{i.e} $\sigma$ deviation, of the sensitive volume ratio with respect to the associated errors, is shown in each bins. For this set the IMR-templates search shows higher sensitivity compared to the ringdown search and the CWB search across the parameter space. The higher sensitivity of the IMR-templates search is significantly more with respect to the ringdown search in comparison with the CWB search. The significance is higher for binary black holes with spin aligned to the angular momentum.
CWB search has more sensitivity compared to Ringdown search for this mass range, albeit within 2 $\sigma$ deviation only.

Fig.~\ref{fig:sensitive-distance-difference-IMR-setB} show the ratio of the sensitive volume for set B injections in total mass, mass ratio and total mass, spin parameter space. Across the parameter space CWB is more sensitive than the ringdown search albeit with varying degree of significance.

%%%%%%%%%%%%%%%%%%%%%%%%%%%%%%%%%%%%%%%%
\section{Conclusion}
This paper introduces a framework for comparing searches for binary black hole coalescences in ground-based gravitational wave detectors, and applies it to  algorithms used in recently published searches~\cite{cbc-highmass,cbc-highmass-s6,imbh,imbh-s6,S5_S6_ringdown} on a 2 months segment of initial LIGO data.
The codes developed for this analysis are available in the LIGO-Scientific Collaboration Algorithm Library software packages~\cite{lal}.

This is the first one-on-one comparison of searches for binary black hole coalescences. The false alarm rate of 3 events per year is a prefatory choice for the demonstration of the method. This analysis provides groundwork for future work  which will include comparison of algorithms at detection-level false alarm rate. 

We provide a quantitative measure of how the algorithms tested in this study were more sensitive to symmetric than asymmetric systems, and to  aligned rather than anti-aligned binary black holes. Additionally, we probe the different sensitivity of the algorithms to different region of the  black hole binary parameter space. 
We find that matched filtering search algorithm using the full Inspiral-Merger-Ringdown templates is the most sensitive search algorithm for total mass between 25 and 100 \msun: 6\% more than a morphology-independent excess power search and  17\% more than matched filtering to ringdown templates, by the measurement of sensitive distance averaged over source parameters at a false alarm rate of 3 events/year. A fully coherent gravitational-wave burst search algorithm is 21\% more sensitive compared to the matched filtering search algorithm with ringdown templates, by the measurement of sensitive distance averaged over source parameters at a false alarm rate of 3 events/year.

This was a non-blind analysis, as the characteristics of simulated signals were known \emph{a priori}. The ability to detect binary black hole coalescence signals through different search algorithms in a blind analysis has been shown in a different study~\cite{ninja-mdc}.
Also, this work does not attempt to combine information from searches. A likelihood based method to combine multiple searches has been prescribed in~\cite{combine-searches}. 

In this study have relied on non-precessing phenomenological waveform model. As new template families with wider coverage of the parameter space are now becoming available~\cite{ninja2catalog,eobnrHM,phenomP},  the method prescribed in this paper can readily be extended to them.
We also note the analysis presented in this paper relies on the initial LIGO sensitivity. We expect differences will ensue with the different possible noise spectra expected in advanced LIGO~\cite{aLIGO-noise,harry-aligo,aligo-imbh} and advanced Virgo~\cite{aVirgo-noise}. Additionally, we do not expect the population of non-Gaussian noise transients to be the same in advanced LIGO/Virgo as in initial LIGO/Virgo data.
New algorithms and pipeline improvements are currently under development; for instance, the inclusion of spinning templates in the bank for IMR-templates~\cite{sbank_highmass}, different clustering in the CWB search~\cite{cwb-doc}, and a more sophisticated post-processing for the ringdown search~\cite{S5_S6_ringdown};  the relative performance of the searches may need to be re-assessed once such developments have stabilized. 

\section{Acknowledgment}
This work was supported by NSF grants PHY-0653550 and PHY-0955773. SM was partly supported by NSF awards  PHY-0855494, PHY-0847611 and PHY-1104371. SC and CP were supported by NSF awards PHY-0970074 and PHY-1307429. SC would like to acknowledge NSF award PHY-0905184. RV and SV were supported by LIGO laboratory. LIGO was constructed by the California Institute of Technology and Massachusetts Institute of Technology with funding from the National Science Foundation and operates under cooperative agreement PHY-0757058. Work on the LIGO-Caltech computing cluster was supported by the NFS grant PHY-0757058. Work on the Nemo cluster was supported by NSF award PHY-1104371. Work on the Sugar cluster was supported by NSF grants PHY-1040231, PHY-1104371 and Syracuse University ITS.

The authors would like to acknowledge Lisa Goggin for the work during the initial phase of this analysis. The authors would like to thank Giulio Mazzolo, Tom Dent, B.S. Sathyaprakash and Stephen Fairhurst for useful discussion and feedback on the manuscript.

\\
\\
This document has been assigned LIGO laboratory document number P1100198.
%%%%%%%%%%%%%%% bib
\\
\\
\bibliographystyle{unsrt}
\bibliography{paper}

\begin{thebibliography}{10}

\bibitem{300years}
S.~Hawking and W.~Israel, editors.
\newblock {\em 300 Years of Gravitation}.
\newblock Cambridge University Press, Cambridge, 1987.

\bibitem{ligo-ref}
B.~P. Abbott et~al.
\newblock {LIGO}: the laser interferometer gravitational-wave observatory.
\newblock {\em Reports on Progress in Physics}, 72(7):076901, 2009.

\bibitem{virgo-ref}
F~Acernese et~al.
\newblock The {V}irgo 3-km interferometer for gravitational wave detection.
\newblock {\em Journal of Optics A: Pure and Applied Optics}, 10(6):064009,
  2008.

\bibitem{polytrope}
S.~L. Shapiro and S.~A. Teukolsky.
\newblock {\em Black Holes, White Dwarfs and Neutron Stars: The Physics of
  Compact Objects}.
\newblock John Wiley and Sons, New York, 1983.

\bibitem{0004-637X-730-2-140}
Tomasz Bulik, Krzysztof Belczynski, and Andrea Prestwich.
\newblock Ic10 x-1/ngc300 x-1: The very immediate progenitors of bh-bh
  binaries.
\newblock {\em The Astrophysical Journal}, 730(2):140, 2011.

\bibitem{2041-8205-715-2-L138}
Krzysztof Belczynski, Michal Dominik, Tomasz Bulik, Richard O’Shaughnessy,
  Chris Fryer, and Daniel~E. Holz.
\newblock The effect of metallicity on the detection prospects for
  gravitational waves.
\newblock {\em The Astrophysical Journal Letters}, 715(2):L138, 2010.

\bibitem{0004-637X-759-1-52}
Michal Dominik, Krzysztof Belczynski, Christopher Fryer, Daniel~E. Holz,
  Emanuele Berti, Tomasz Bulik, Ilya Mandel, and Richard O'Shaughnessy.
\newblock Double compact objects. {I}. the significance of the common envelope
  on merger rates.
\newblock {\em The Astrophysical Journal}, 759(1):52, 2012.

\bibitem{imbh-globular-1}
M.~Coleman Miller and Douglas~P. Hamilton.
\newblock {Production of intermediate-mass black holes in globular clusters}.
\newblock {\em Mon.Not.Roy.Astron.Soc.}, 330:232, 2002.

\bibitem{imbh-globular-2}
Thomas~J. Maccarone, Arunav Kundu, Stephen~E. Zepf, and Katherine~L. Rhode.
\newblock {A black hole in a globular cluster}.
\newblock {\em Nature}, 445:183--185, 2007.

\bibitem{0004-637X-734-2-111}
Shane~W. Davis, Ramesh Narayan, Yucong Zhu, Didier Barret, Sean~A. Farrell,
  Olivier Godet, Mathieu Servillat, and Natalie~A. Webb.
\newblock The cool accretion disk in eso 243-49 hlx-1: Further evidence of an
  intermediate-mass black hole.
\newblock {\em The Astrophysical Journal}, 734(2):111, 2011.

\bibitem{inspiral}
B.~P. Abbott et~al.
\newblock {Search for Gravitational Waves from Low Mass Binary Coalescences in
  the First Year of \uppercase{LIGO}$'$s S5 Data}.
\newblock {\em Phys. Rev.}, D79:122001, 2009.

\bibitem{cbclowmass_s6}
J.~Abadie et~al.
\newblock Search for gravitational waves from low mass compact binary
  coalescence in {LIGO’}s sixth science run and virgo’s science runs 2 and
  3.
\newblock {\em Phys. Rev. D}, 85:082002, Apr 2012.

\bibitem{cbc-highmass}
J.~Abadie et~al.
\newblock Search for gravitational waves from binary black hole inspiral,
  merger, and ringdown.
\newblock {\em Phys. Rev. D}, 83(12):122005, Jun 2011.

\bibitem{cbc-highmass-s6}
J.~Aasi et~al.
\newblock Search for gravitational waves from binary black hole inspiral,
  merger, and ringdown in {LIGO}-{V}irgo data from 2009-2010.
\newblock {\em Phys. Rev. D}, 87:022002, Jan 2013.

\bibitem{S5_S6_ringdown}
J.~Aasi et~al.
\newblock {Search for gravitational wave ringdowns from perturbed intermediate
  mass black holes in {LIGO}-{V}irgo data from 2005-2010}.
\newblock 2014.

\bibitem{imbh}
J.~Abadie et~al.
\newblock Search for gravitational waves from intermediate mass binary black
  holes.
\newblock {\em Phys. Rev. D}, 85:102004, May 2012.

\bibitem{imbh-s6}
Aasi et~al.
\newblock Search for gravitational radiation from intermediate mass black hole
  binaries in data from the second {LIGO}-{V}irgo joint scientific run.
\newblock {\em In Preparation}, 2014.

\bibitem{adv-LIGO}
Advanced {LIGO} -- {T}he {N}ext {S}tep in {G}ravitational {W}ave {A}stronomy.
\newblock \url{https://www.advancedligo.mit.edu/}.

\bibitem{adv-Virgo}
{I}ntroduction to {A}dvance {V}irgo.
\newblock \url{https://wwwcascina.virgo.infn.it/advirgo/}.

\bibitem{rate}
J.~Abadie et~al.
\newblock Predictions for the rates of compact binary coalescences observable
  by ground-based gravitational-wave detectors.
\newblock {\em Classical and Quantum Gravity}, 27(17):173001, 2010.

\bibitem{flanhughes}
\'E.~\'E. Flanagan and S.~A. Hughes.
\newblock Measuring gravitational waves from binary black hole coalescences.
  {I}. signal to noise for inspiral, merger, and ringdown.
\newblock {\em Phys. Rev. D}, 57(8):4535--4565, Apr 1998.

\bibitem{ringdown}
B.~P. Abbott et~al.
\newblock {Search for gravitational wave ringdowns from perturbed black holes
  in \uppercase{LIGO} S4 data}.
\newblock {\em Phys. Rev.}, D80:062001, 2009.

\bibitem{chi-square-bruce}
Bruce Allen.
\newblock Chi-square time-frequency discriminator for gravitational wave
  detection.
\newblock {\em Phys. Rev. D}, 71:062001, Mar 2005.

\bibitem{multi-resolution}
S~Chatterji, L~Blackburn, G~Martin, and E~Katsavounidis.
\newblock Multiresolution techniques for the detection of gravitational-wave
  bursts.
\newblock {\em Classical and Quantum Gravity}, 21(20):S1809, 2004.

\bibitem{robinson2008}
C.~A.~K. Robinson, B.~S. Sathyaprakash, and Anand~S. Sengupta.
\newblock Geometric algorithm for efficient coincident detection of
  gravitational waves.
\newblock {\em Phys. Rev. D}, 78:062002, Sep 2008.

\bibitem{cwb}
S~Klimenko, I~Yakushin, A~Mercer, and G~Mitselmakher.
\newblock A coherent method for detection of gravitational wave bursts.
\newblock {\em Classical and Quantum Gravity}, 25(11):114029, 2008.

\bibitem{s5-sensitivity}
J.~Abadie et~al.
\newblock {Sensitivity to Gravitational Waves from Compact Binary Coalescences
  Achieved during {LIGO}$'$s Fifth and {V}irgo$'$s First Science Run}.
\newblock 2010.

\bibitem{ligo-wa}
{LIGO} {H}anford {O}bservatory.
\newblock \url{http://www.ligo-wa.caltech.edu/}.

\bibitem{ligo-la}
{LIGO} {L}ivingston {O}bservatory.
\newblock \url{http://www.ligo-la.caltech.edu/}.

\bibitem{calibration}
J.~Abadie et~al.
\newblock Calibration of the {LIGO} gravitational wave detectors in the fifth
  science run.
\newblock {\em Nuclear Instruments and Methods in Physics Research Section A:
  Accelerators, Spectrometers, Detectors and Associated Equipment}, 624(1):223
  -- 240, 2010.

\bibitem{s5-glitch}
L~Blackburn et~al.
\newblock The {LSC} glitch group: monitoring noise transients during the fifth
  {LIGO} science run.
\newblock {\em Classical and Quantum Gravity}, 25(18):184004, 2008.

\bibitem{first_joint_ligo_geo_virgo_burst_s5}
J.~Abadie et~al.
\newblock All-sky search for gravitational-wave bursts in the first joint
  {LIGO}-{GEO}-{V}irgo run.
\newblock {\em Phys. Rev. D}, 81:102001, May 2010.

\bibitem{eobnr}
A.~Buonanno et~al.
\newblock Approaching faithful templates for nonspinning binary black holes
  using the effective-one-body approach.
\newblock {\em Physical Review D (Particles, Fields, Gravitation, and
  Cosmology)}, 76(10):104049, 2007.

\bibitem{Teukolsky-1973ha}
Saul~A. Teukolsky.
\newblock {Perturbations of a rotating black hole. 1. Fundamental equations for
  gravitational electromagnetic and neutrino field perturbations}.
\newblock {\em Astrophys.J.}, 185:635--647, 1973.

\bibitem{creighton-ring-template}
Jolien D.~E. Creighton.
\newblock Search techniques for gravitational waves from black-hole ringdowns.
\newblock {\em Phys. Rev. D}, 60:022001, Jun 1999.

\bibitem{berti_2006}
Emanuele Berti, Vitor Cardoso, and Clifford~M. Will.
\newblock Gravitational-wave spectroscopy of massive black holes with the space
  interferometer lisa.
\newblock {\em Phys. Rev. D}, 73:064030, Mar 2006.

\bibitem{nakano2003}
Hiroyuki Nakano, Hirotaka Takahashi, Hideyuki Tagoshi, and Misao Sasaki.
\newblock An effective search method for gravitational ringing of black holes.
\newblock {\em Phys. Rev. D}, 68:102003, Nov 2003.

\bibitem{talukder2003}
Dipongkar Talukder, Sukanta Bose, Sarah Caudill, and Paul~T. Baker.
\newblock Improved coincident and coherent detection statistics for searches
  for gravitational wave ringdown signals.
\newblock {\em Phys. Rev. D}, 88:122002, Dec 2013.

\bibitem{second_joint_ligo_virgo_burst_s6}
J.~Abadie et~al.
\newblock All-sky search for gravitational-wave bursts in the second joint
  {LIGO}-{V}irgo run.
\newblock {\em Phys. Rev. D}, 85:122007, Jun 2012.

\bibitem{elliptical-polarization}
C~Pankow, S~Klimenko, G~Mitselmakher, I~Yakushin, G~Vedovato, M~Drago, R~A
  Mercer, and P~Ajith.
\newblock A burst search for gravitational waves from binary black holes.
\newblock {\em Classical and Quantum Gravity}, 26(20):204004, 2009.

\bibitem{spin-phenom}
P.~Ajith, M.~Hannam, S.~Husa, Y.~Chen, B.~Br\"ugmann, N.~Dorband, D.~M\"uller,
  F.~Ohme, D.~Pollney, C.~Reisswig, L.~Santamar\'\i{}a, and J.~Seiler.
\newblock Inspiral-merger-ringdown waveforms for black-hole binaries with
  nonprecessing spins.
\newblock {\em Phys. Rev. Lett.}, 106(24):241101, Jun 2011.

\bibitem{orbital-hangup}
Manuela Campanelli, C.O. Lousto, and Y.~Zlochower.
\newblock {Spinning-black-hole binaries: The orbital hang up}.
\newblock {\em Phys.Rev.}, D74:041501, 2006.

\bibitem{final-spin}
Luciano Rezzolla, Peter Diener, Ernst~Nils Dorband, Denis Pollney, Christian
  Reisswig, Erik Schnetter, and Jennifer Seiler.
\newblock The final spin from the coalescence of aligned-spin black hole
  binaries.
\newblock {\em The Astrophysical Journal Letters}, 674(1):L29, 2008.

\bibitem{obs-scenario}
J.~Aasi et~al.
\newblock {Prospects for {L}ocalization of {G}ravitational {W}ave {T}ransients
  by the {A}dvanced {LIGO} and {A}dvanced {V}irgo Observatories}.
\newblock 2013.

\bibitem{0264-9381-24-12-S04}
Frank Herrmann, Ian Hinder, Deirdre Shoemaker, and Pablo Laguna.
\newblock Unequal mass binary black hole plunges and gravitational recoil.
\newblock {\em Classical and Quantum Gravity}, 24(12):S33, 2007.

\bibitem{loudest1}
Rahul Biswas, Patrick~R Brady, Jolien D~E Creighton, and Stephen Fairhurst.
\newblock The loudest event statistic: general formulation, properties and
  applications.
\newblock {\em Classical and Quantum Gravity}, 26(17):175009, 2009.

\bibitem{loudest2}
Patrick~R Brady and Stephen Fairhurst.
\newblock Interpreting the results of searches for gravitational waves from
  coalescing binaries.
\newblock {\em Classical and Quantum Gravity}, 25(10):105002, 2008.

\bibitem{science-summary}
Comparing search techniques for studying collisions of binary black holes.
\newblock
  \url{https://www.gravity.phy.syr.edu/~satya/Publication-IMR_comparison/},
  2014.

\bibitem{ninja-mdc}
J.~Aasi, , et~al.
\newblock The {NINJA}-2 project: detecting and characterizing gravitational
  waveforms modelled using numerical binary black hole simulations.
\newblock {\em Classical and Quantum Gravity}, 31(11):115004, 2014.

\bibitem{lal}
LIGO~Scientific Collaboration.
\newblock \uppercase{LSC} algorithm library software packages \uppercase{lal},
  \uppercase{lalwrapper}, and \uppercase{lalappsl}.
\newblock \url{http://www.lsc-group.phys.uwm.edu/lal}.

\bibitem{combine-searches}
Rahul Biswas, Patrick~R. Brady, Jordi Burguet-Castell, Kipp Cannon, Jessica
  Clayton, Alexander Dietz, Nickolas Fotopoulos, Lisa~M. Goggin, Drew Keppel,
  Chris Pankow, Larry~R. Price, and Ruslan Vaulin.
\newblock Detecting transient gravitational waves in non-gaussian noise with
  partially redundant analysis methods.
\newblock {\em Phys. Rev. D}, 85:122009, Jun 2012.

\bibitem{ninja2catalog}
P.~Ajith et~al.
\newblock {The NINJA-2 catalog of hybrid post-Newtonian/numerical-relativity
  waveforms for non-precessing black-hole binaries}.
\newblock {\em Class. Quant. Grav.}, 29(12):124001, 2012.

\bibitem{eobnrHM}
Andrea Taracchini, Yi~Pan, Alessandra Buonanno, Enrico Barausse, Michael Boyle,
  Tony Chu, Geoffrey Lovelace, Harald~P. Pfeiffer, and Mark~A. Scheel.
\newblock Prototype effective-one-body model for nonprecessing spinning
  inspiral-merger-ringdown waveforms.
\newblock {\em Phys. Rev. D}, 86:024011, Jul 2012.

\bibitem{phenomP}
Mark Hannam, Patricia Schmidt, Alejandro Bohé, Leila Haegel, Sascha Husa,
  et~al.
\newblock {{T}wist and shout: {A} simple model of complete precessing
  black-hole-binary gravitational waveforms}.
\newblock 2013.

\bibitem{aLIGO-noise}
Advanced {LIGO} anticipated sensitivity curves.
\newblock \url{https://dcc.ligo.org/LIGO-T0900288/public}.

\bibitem{harry-aligo}
Gregory~M Harry and the LIGO Scientific~Collaboration.
\newblock Advanced {LIGO}: the next generation of gravitational wave detectors.
\newblock {\em Classical and Quantum Gravity}, 27(8):084006, 2010.

\bibitem{aligo-imbh}
Giulio Mazzolo et~al.
\newblock Prospects for intermediate mass black hole binary searches with
  advanced gravitational-wave detectors.
\newblock \url{https://dcc.ligo.org/P1300053}.

\bibitem{aVirgo-noise}
Advanced {V}irgo {B}aseline {D}esign.
\newblock
  \url{https://tds.ego-gw.it/itf/tds/file.php?callFile=VIR-0027A-09.pdf}.

\bibitem{sbank_highmass}
Stephen Privitera, Satya Mohapatra, Parameswaran Ajith, Kipp Cannon, Nickolas
  Fotopoulos, Melissa~A. Frei, Chad Hanna, Alan~J. Weinstein, and John~T.
  Whelan.
\newblock Improving the sensitivity of a search for coalescing binary black
  holes with nonprecessing spins in gravitational wave data.
\newblock {\em Phys. Rev. D}, 89:024003, Jan 2014.

\bibitem{cwb-doc}
{CWB} {P}ipeline {O}nline {D}ocumentation.
\newblock
  \url{https://atlas3.atlas.aei.uni-hannover.de/~waveburst/LSC/doc/cwb/man/}.

\end{thebibliography}
\end{document}